# Tuneable optical gain and broadband lasing driven in electrospun polymer fibers by high dye concentration

*Giovanni Morello,[a] Maria Moffa,[b] Martina Montinaro,[a] Annachiara Albanese,[b,c] Karolis Kazlauskas,[d] Saulius Jursenas,[d] Ausra Tomkeviciene,[e] Juozas V. Grazulevicius,[e] Andrea Camposeo[b,†] and Dario Pisignano[a,b,c,†]*

[a] Dipartimento di Matematica e Fisica "Ennio De Giorgi", Università del Salento, via Arnesano I-73100 Lecce, Italy.

[b] NEST, Istituto Nanoscienze-CNR and Scuola Normale Superiore, Piazza S. Silvestro 12, I-56127 Pisa, Italy.

[c] Dipartimento di Fisica, Università di Pisa, Largo B. Pontecorvo 3, I-56127 Pisa, Italy.

[d] Institute of Photonics and Nanotechnology, Vilnius University, Sauletekio Av. 3, LT-10257, Vilnius, Lithuania.

[e] Department of Polymer Chemistry and Technology, Kaunas University of Technology, Radvilenu pl. 19, LT-50254 Kaunas, Lithuania

[†] Corresponding authors e-mail: andrea.camposeo@nano.cnr.it; dario.pisignano@unipi.it.

**Abstract**

The optical gain of blue light-emitting electrospun polystyrene fibers doped with a linear multi-fragment molecular dye based on the combination of fluorine-carbazole functional units is investigated, with the aim of correlating emission properties and the specific material architecture made of either aligned or disordered fibers. Enhanced performance is found in aligned fibers, whose gain spectrum can be finely tuned by varying the dye concentration. Instead, randomly oriented fibers show a manifold spectral line narrowing, resulting in sharp laser peaks superimposed on top of a broad emission band, ascribable to random lasing. In these systems, the increase of the dye content turns out to be effective for both decreasing the lasing threshold by about a factor 6 and for varying the laser emission wavelength. These results make these arrays and disordered architectures of fibers valuable active media for variable-gain, broadband lasing, which is remarkably important in optical sensing and tunable microlaser devices.

**1. Introduction**

The achievement of broadband and random lasing (RL) in disordered materials[1–3] has fuelled many research efforts aimed at understanding the basic physical mechanisms of lasing in complex media and the specific properties of such lasers, which might include low spatial coherence, multimode emission, and either highly correlated or chaotic emission. These properties can be exploited in novel applications, such as speckle-free imaging,[4] white lasing,[5] and high-resolution spectroscopy.[6] Among the complex media that can simultaneously feature optical gain, light diffusion and either multimode or broadband emission, polymer fibers are emerging as promising building blocks for a new generation of optical components, such as micro-sized light sources and amplifiers,[7–9] lasers[10,11] and optical sensors.[12,13] Polymer fibers can be realized by various fabrication methods,[14,15] among which electrospinning[16] is highly advantageous in terms of throughput and capability to control the optical properties of the individual filaments and of the complex networks made of them. Indeed, diverse architectures can be electrospun,[17–19] including aligned arrays, periodic patterns and randomly distributed fibers that all are appealing for application in photonics. For instance, randomly distributed fibers are characterized by a high degree of disorder and spatial variability of both local fiber density and orientation, leading to the possibility to engineer the effects of disorder through the specific architecture and topology of the network formed.[20] The study of lasing from disordered polymer fibers received an increasing attention in recent years, and several theoretical and experimental works have been performed to understand the origin and the possible spatial and temporal evolution of the observed lasing modes.[10,20–25] These studies rationalize the activation of lasing in localized domains characterized by an ensemble of fiber segments that guide light, and a number of scattering sites at both the fiber intersection points[20] and eventually embedded nanoparticles.[21,25] In this framework, recent evidences support the high degree of correlation among laser peaks in shot-to-shot spectra[21,24] and the possibility to observe spectral and spatial replica symmetry breaking processes,[26] revealing RL switching occurring inside the same excited region and involving neighbour spatial domains extended over a few tens of microns.[23] Dependencies on the excited areas and pumping level have been also evidenced.[20,23] The achievement of tunability of the emission constitutes another relevant issue, especially for sensing, medical diagnostic and optical communication applications. At variance with ordinary lasing systems in which a number of strategies are employed to control the cavity resonance wavelengths involving the variation of the resonator geometry,[27,28] in RL systems more complex methods are used to achieve spectral control, based for instance on micro-scatterers (spheres) embedded in gain medium[29] and stretchable materials.[30] A number of strategies for effective control and tuning of the emission of fibers array have been reported, exploiting the variation of the diameter of individual fiber or loops made of them, of the refractive index of the fibers, of their surrounding environment, or of other ambient parameters including temperature.[12,13,31–33] Reaching effective tunability of the emission of RL fibers would require both broadband optical gain and high net gain coefficients, especially when the gain medium is miniaturized down to the scale of either individual fibers or planar fiber networks, with optical losses becoming significant and potentially preventing low-threshold lasing. In general, strategies to enhance the optical gain of molecular systems might include the use of blends of donor–acceptor conjugated polymer couples,[34] complexes as triplet sensitizers,[35] or extended ladder-type oligomers.[36] In addition, increasing the content of the embedded gain nanoparticles or molecules constitutes an effective route for enhancing the net gain, though limited by critical concentration values leading to dopant aggregation (with associated increase of scattering losses and suppressed dye emission).

In this work, the optical gain of polymer fibers incorporating a blue-emitting chromophore is studied, and used as a bench tool for analysing the possibility to finely tune lasing through dye concentration while preserving good gain performance. The gain properties of different architectures and composition of the fibers are compared. Arrays of uniaxially aligned fibers display high net optical gain and low losses, whereas random ones show broadband RL with excitation threshold effectively reduced by increasing the amount of incorporated fluorescent molecules. In both the fibrous architectures, the peak wavelength of the gain and the lasing features a well-defined dependence on the dye concentration. These results might establish new design guidelines and provide further versatility for the realization and development of broadband tuneable laser sources based on fibrous architectures.

## 2. Results and discussion

The chromophore used for light emission and amplification is a multi-fragment molecular dye, 2,7-bis(9,9-diethylfluoren-2-yl)-9-(2-ethylhexyl)carbazole (hereafter denoted as FCF), based on the combination of fluorine and carbazole functional units (molecular structure in Fig. 1a). The synthesis procedure and optical properties in thin films of polystyrene (PS) have been reported previously.[37,38] FCF has shown to exhibit relatively weak concentration quenching due to the twisted molecular backbone and the presence of bulky ethyl and ethylhexyl groups.[38]

Electrospun fibers (Fig. 1b and c) are made by embedding FCF in a PS matrix, here selected for the high optical transparency in the visible range.[39] By varying the deposition conditions, it is possible to obtain bundles of randomly oriented and uniaxially aligned fibers (see the Experimental section). The resulting fibers emit blue light and show bright, uniform and defect-free emission along their length as evidenced by confocal fluorescence microscopy (Fig. 1b), indicative of a homogeneous distribution of the dye and of the lack of appreciable clustering effects. The electrospun fibers have an average diameter of 4 μm, as measured by scanning electron microscopy (SEM, Fig. 1c).

Fibers are realized by five different concentrations, varying the FCF/polymer weight ratio ($\chi$) in the range 0.5–10%. It is worth to note that the different FCF/polymer weight ratios do not modify the fiber morphology.

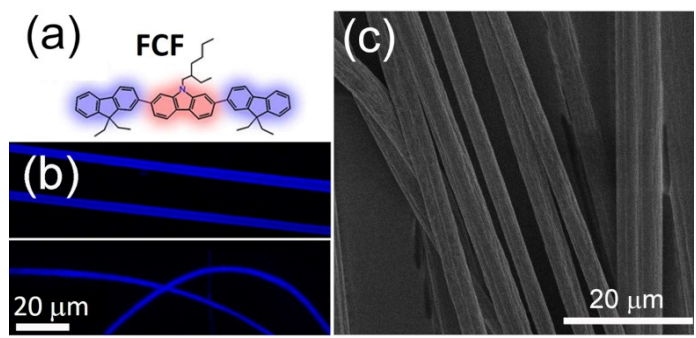

**Figure 1.** (a) FCF molecular structure. (b) Confocal fluorescence micrographs of aligned (top image) and randomly oriented (bottom image) fibers. (c) SEM micrograph of uniaxially aligned fibers.

First, the emission properties of the uniaxially aligned fibers are characterized, which are prone to light amplification due to the self-waveguiding of the light emitted by FCF molecules along the fiber length, as observed in various classes of electrospun fibers.[7,9,40] The amplified spontaneous emission (ASE) and gain properties of aligned fibers with FCF concentration of 5% are shown in Fig. 2. ASE is analyzed by the Light-in/Light-out (L–L) dependence, obtained by optically pumping the fibers with a stripe-shaped beam positioned with its length parallel to the fiber alignment direction and with one end placed in correspondence of the fibers emitting edge. As expected for materials showing ASE, at lower excitation levels the spectra consist mainly of spontaneous emission, whereas as long as the pumping intensity is increased, the full width at half maximum (FWHM) of the spectra is decreased as a result of the light amplification process (Fig. 2a and b). ASE is initiated by spontaneous emission which is then amplified through stimulated emission because of the population inversion achieved by optical pumping.[41] Though ASE is a thresholdless phenomenon, a conventional threshold can be calculated as the fluence at which the FWHM takes the halfway value between the maximum and the minimum. This estimate leads to a threshold of 18 mJ cm$^{-2}$ for the uniaxially aligned fibers (Fig. 2). Since the optical gain is directly related to the extension of the light path inside the region undergoing population inversion (*i.e.* the size of the photo-excited region), it can be measured by exciting the fibers at a fixed fluence and measuring the emission spectra by varying the excitation stripe length ($L$). At a specific wavelength, the emission intensity ($I_L$) is related to the net optical gain, $G(\lambda)$, experienced by the material by means of the following expression:[42]

$$I_L = \frac{I_p A(\lambda)}{G(\lambda)} \cdot \left[ e^{G(\lambda) \cdot L} - 1 \right] \quad (1)$$

where $I_p$ is the pump intensity and $A(\lambda)$ is a factor accounting for the spontaneous emission cross-section. The spectrum of the net optical gain of the fibers gives insight about the peak gain wavelength and on bandwidth. By fitting the experimental data to Eq. (1) (an example of this analysis is reported in the Supplementary Information, SI, Fig. S1) the gain spectrum is obtained, as shown in Fig. 2c.

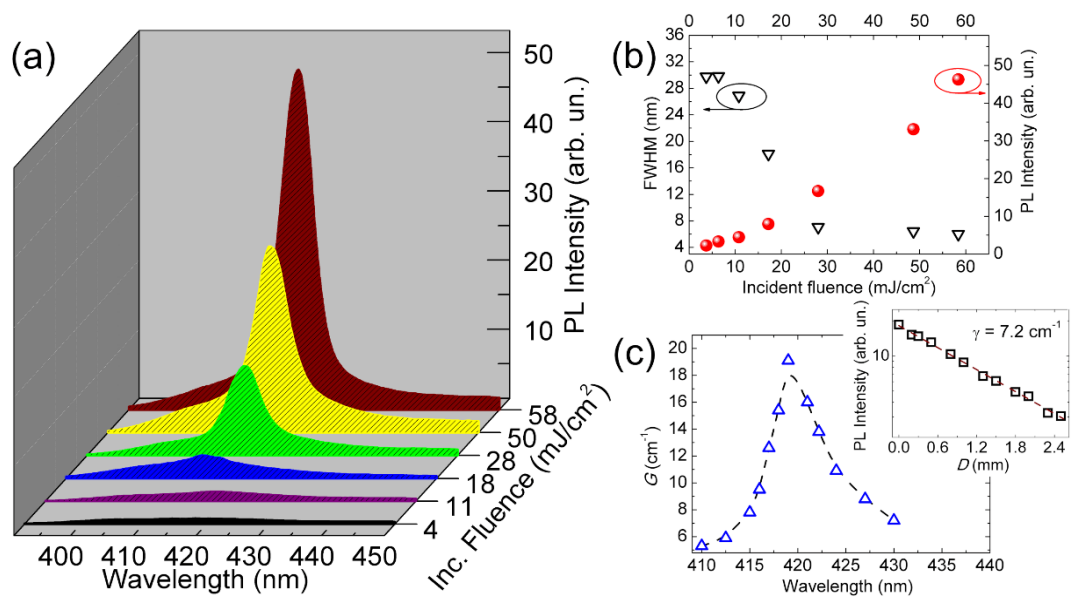

**Figure 2.** (a) ASE spectra from aligned FCF-doped PS fibers ($\chi = 5\%$) for various incident fluences. (b) Emission intensity (full symbols, right vertical scale) and FWHM (empty symbols, left vertical scale) as a function of the incident fluence. (c) Optical gain spectrum of the same sample. Each gain value is obtained by analysing the ASE intensity at a certain wavelength as a function of the stripe length and fitted to Eq. (1). The dashed line is a guide for the eyes. Inset: Plot of the ASE intensity *vs.* distance from the sample edge. The dashed line is a fit to the data by an exponential function.

Furthermore, optical loss coefficients ($\gamma$) can be measured by analyzing the emission intensity ($I_{PL}$) as a function of the distance, $D$, of the excitation stripe from the emitting edge of the fibers, by means of the expression $I_{PL} = I_0 e^{-\gamma D}$. The inset of Fig. 2c reports the analysis for aligned fibers with $\chi = 5\%$, performed by integrating the ASE intensity in the spectral interval 416–423 nm (*i.e.*, around the wavelength of the maximum optical gain), which shows low optical propagation losses, as also found in various other dye-doped polymer systems[8,9,40] and typically ascribed to self-absorption, and light scattering by surface and bulk defects.[43] These losses constitute the most disadvantageous mechanism hindering efficient optical gain in self-waveguiding active media.[40] They can be tackled by both the choice of the dye and its concentration. For instance, FCF exhibits a large Stokes shift (480 meV, Fig. S2 in the SI), leading to a decrease of the self-absorption contribution.[8,9] Studying the gain parameters at variable dye concentration is essential for defining the more suitable conditions for the lowest excitation threshold and losses, and the highest gain. While the increase of $\chi$ is expected to enhance the gain capability (low threshold and high net gain), an excessive amount of dye incorporated into the fibers might lead to a degradation of the optical performance due to concentration-induced quenching of the luminescence.[38] Fig. 3a shows the ASE spectra of aligned fibers relative to the five investigated weight ratios and excited above threshold, whereas Fig. 3b shows the dependence of the maximum of the net gain, $G_{max}$, of the losses coefficient and of the ASE peak wavelength on $\chi$. A noticeable red shift is observed upon increasing $\chi$, a behavior that can be ascribed to the absorption losses (increasing with $\chi$), as already observed in other light-emitting systems.[44] In fact, upon increasing $\chi$ the probability of self-absorption losses for the spontaneous emission waveguided along a fiber is increased, especially for those photons emitted at the shorter wavelengths of the PL spectrum, because of the higher overlap with the low-energy tail of the absorption band (Fig. S2, SI). This enhances the propagation losses for photons emitted at shorter wavelengths, thus preventing their amplification through stimulated emission and leading to a red-shift of net optical gain. Moreover, the maximum net gain shows a significant increase for higher values of $\chi$ in the interval 0.5–2%, and it keeps almost constant upon further increasing of the dye amount (Fig. 3b). Such values are higher than net gain typically reported for dye-doped polymer fibers (see Table S1 of the SI) emitting in the visible spectral range, and in line with the best gain performance reported for red-emitting polymer fibers doped by 4-(dicyanomethylene)-2-*tert*-butyl-6(1,1,7,7-tetramethyljulolidyl-9-enyl)-4*H*-pyran.[45,46]

The observed trends can be rationalized taking into account that while high values of $\chi$ can be in principle effective for enhancing the gain and lowering the ASE threshold, the increase of the FCF concentration can, on the other hand, favor the decrease of the luminescence quantum efficiency and the gain *via* quenching mechanisms, as reported for the same molecular system dispersed in PS films.[38] By analyzing the behavior of $\gamma$, an increase is measured up to $\chi = 5\%$, followed by minor changes for relative concentration values up to 10%. It is largely reported that the self-absorption plays a crucial role in defining the optical losses in a waveguide and that the higher is the dye concentration, the higher is self-absorption.[9,40]

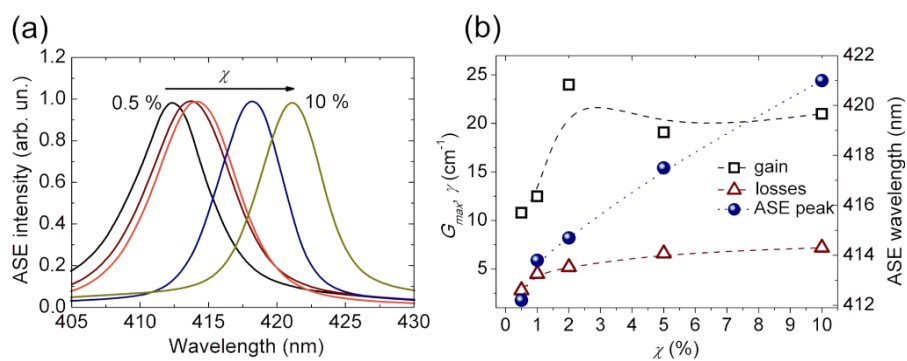

**Figure 3.** (a) ASE from aligned FCF-doped PS fibers with varied dye concentration. (b) Dependence of the maximum net gain (squares, left vertical scale), losses (triangles, left vertical scale) and ASE peak wavelength (circles, right vertical scale) vs. $\chi$ for uniaxially aligned fibers. The dashed lines are guides for the eyes.

This implies that an excess amount of self-absorbing active material induces a monotonic increase of $\gamma$. Overall, data of Fig. 3 clearly show that the use of the FCF dye allows the spectral properties of the gain band of the electrospun fibers to be tailored, keeping almost constant the net gain at high concentrations. This determines a well-defined critical value for $\chi$ (Fig. 4), above which the quenching of emission intensity overcomes the increase of net gain with a resulting increase of the threshold. Such limit of $\chi$ is in the range 2–5% for the uniaxially aligned fibers (minimum measured ASE threshold = 15 mJ cm$^{-2}$ at $\chi$ = 2%). These concentrations are among the highest reported for dye-doped fibers (see Table S1 in the SI) making FCF an ideal molecular dye for an effective variation of the gain band.

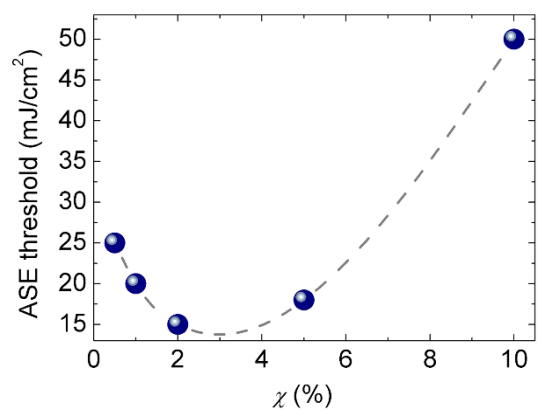

**Figure 4.** Excitation threshold vs. FCF weight ratio, $\chi$, for aligned fibers. Dashed line is a guide for the eyes.

Fig. 5a–c shows the morphology of randomly oriented electrospun fibers. These fibers feature surface roughness (inset of Fig. 5c), which is not appreciable in uniaxially aligned ones (Fig. 1c). The roughness suppression effect can be in part associated to the stretching induced by the rotating collector used to obtain aligned fibers. The dependence of the emission spectra on the excitation fluence is shown in Fig. 5d. Line narrowing is found at fluences higher than 50 mJ cm$^{-2}$, and the FWHM decreases to values around 10 nm while the overall intensity increases almost linearly (Fig. 5e). A remarkable property of random fibers is the appearance of sharp peaks on top of a broadband emission spectrum (Fig. 5d), that could be ascribed to a dominant incoherent feedback mechanism with the addition of initial coherent feedback occurring inside the network.[9] Indeed, considering that aligned fibers do not show any feature attributable to genuinely lasing mechanisms, the RL peaks shown in Fig. 5d should be related to the complex network formed by the randomly oriented electrospun fibers. Such features can be further enhanced by varying the degree of complexity and disorder of the fiber network, by adding $TiO_2$ nanoparticles in the fibers (Fig. S3, SI).[21,22] The resulting morphological and compositional properties are summarized in Fig. 6, whereas the main optical properties are reported in Fig. S4 (SI). It is worth to note that $TiO_2$ nanoparticles can show a tendency to aggregate in solution (light scattering reveals an average cluster diameter of 70 nm, with a dispersion of 12 nm), and larger clusters might form upon solvent evaporation (Fig. S3, SI). Fig. 6a and b show the scanning transmission electron microscopy (STEM) analysis performed on single fibers added with $TiO_2$. The particle size inside the fibers can vary from the sub-micron scale to a few microns. Fig. 6b shows an energy dispersive X-ray (EDX) map of a segment of a fiber embedding a large $TiO_2$ cluster, whereas Fig. 6c reports the EDX signal as recorded from two different regions of the fiber (in and out of the $TiO_2$-rich region).

The addition of $TiO_2$ is known to increase the number of scattering centers and the out-coupling efficiency of the light emitted by the FCF molecules, while it decreases the mean path of light and the waveguiding along the fiber length, with the occurrence of well-defined, narrow lines (linewidth comparable to the spectral resolution of the spectrometer, 0.3 nm) in a bandwidth of about 10 nm (Fig. S4b, SI). Upon varying the excitation fluence, the position of these spectral lines is highly stable within the spectral resolution of the spectrometer.

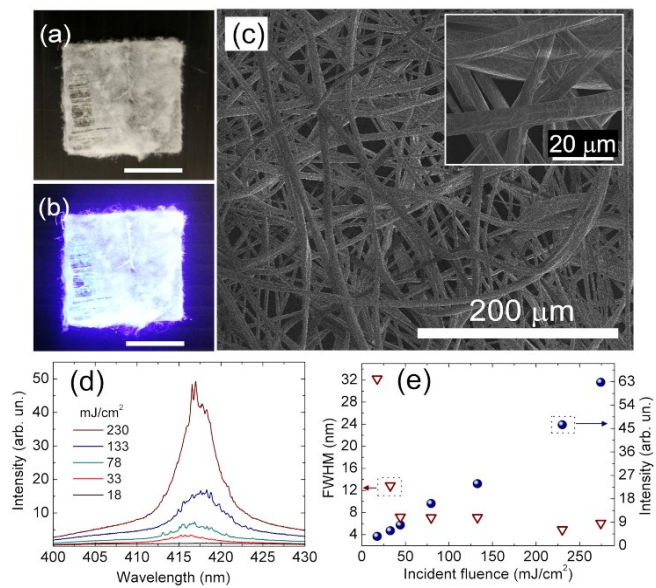

**Figure 5.** Properties of randomly oriented fibers with $\chi = 5\%$. (a and b) Photographs of fibers deposited on a quartz substrate under ambient (a) and UV (b) light. Scale bars: 1 cm. (c) SEM micrographs of dense randomly oriented fibers. Inset: Magnified SEM micrograph. (d) PL spectra measured at different incident fluences. (e) L–L plot for randomly-oriented fibers (full dots, right vertical scale). The corresponding dependence of the FWHM of the spectra *vs.* incident fluence is also shown (empty triangles, left vertical scale).

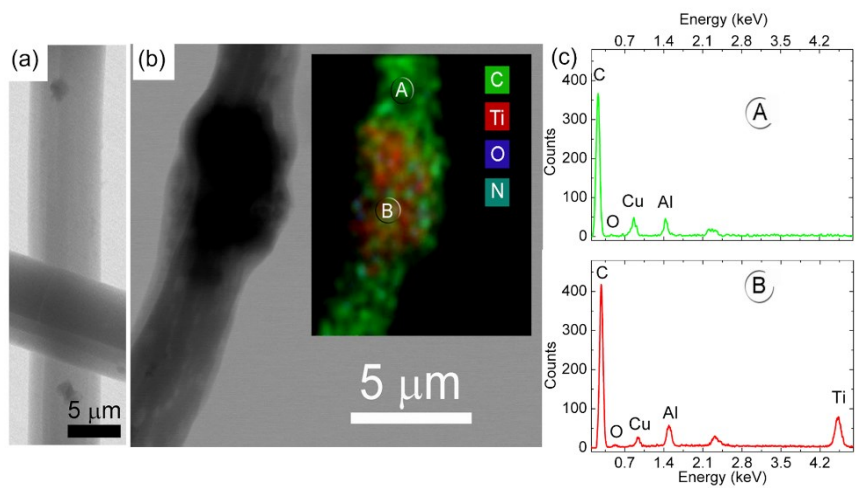

**Figure 6.** (a) STEM micrograph of fiber segments doped with $TiO_2$, showing clusters of $TiO_2$ particles distributed inside the fibers. (b) STEM micrograph and relative elemental analysis by EDX (inset) of a fiber segment. (c) EDX spectra taken at the different regions on the fiber, as indicated by lettering (A) and (B). The noticeable EDX peak of Titanium features the spectrum from (B) with respect to (A).

The L–L plot of two representative peaks (412 nm and 417.5 nm) is shown in Fig. S4c (SI). Both peaks show a linear power-dependent intensity above a well-defined threshold (about 12 mJ cm$^{-2}$), typical of lasing emission.

Regarding the origin of the lasing peaks in randomly oriented fibers (Fig. 5), light-scattering enhanced by FCF aggregation can be ruled out because it should cause the occurrence of RL also in aligned fibers, especially at high dye concentrations. RL from active polymer fibers has been observed in a number of different geometries and networks.[20–24,33,47,48] An accepted scenario involves the localization of lasing modes across a number of possible resonance configurations.[2] Models based on quantum graphs of 2D networks of fibers with diameters comparable to the wavelength of the emission predicted the existence of modes as a result of multi-path interference processes, sustained by the establishment of numerous different closed paths and the localization of the lasing modes as a function of the degree of randomness (number of guiding segments and scattering nodes).[20] Pearson's correlation analyses performed on shot-to-shot RL spectra showed the actual role of a few resonance dynamics resulting from selective interference.[21–24]

The modes guided for fibers with cylindrical geometry as a function of the fiber diameter, $D_F$, for light at 417 nm, *i.e.* where maximum gain of the FCF molecules occurs, are shown in Fig. S5 (SI), where the propagation constant, $\beta$, for the first ten modes is shown. The spatial mode profiles of a few representative modes are also shown in Fig. S5 (SI).

For fibers with size comparable to the one shown in Fig. 5 ($D_F/\lambda \sim 9.6$) a large number of modes can propagate through the fibers, even though the one having better overlap with the gain spatial profile (*i.e.* the FCF molecules embedded into the fibers) is the fundamental one. One can expect that given the average size of the fiber the fundamental mode is almost confined in the fiber body (the fractional mode power of the fundamental mode confined within the fiber is close to 1[49]) and the emission properties are mainly determined by individual fibers.

Indeed, in uniaxially aligned fibers, where light can propagate through the fiber with minimal losses, ASE is mainly observed, whereas in randomly oriented fibers some inter-fiber coupling can occur at the fiber junctions, likely enhanced by the observed surface roughness, determining the emergence of some lasing peaks at high pumping fluences. These effects are also expected to increase propagation losses and to decrease the net gain in randomly oriented fibers, which indeed feature a significantly higher excitation threshold with respect to the aligned ones (Fig. 2b and 5e). Notably, in randomly oriented fibers the lasing peaks are progressively red-shifted upon increasing the FCF/PS weight ratio, $\chi$, with an overall variation of the emission wavelength of about 10 nm (Fig. 7). Here the increase of the dye amount determines also a significant decrease of the lasing threshold, with a minimum value achieved at $\chi = 5\%$ that is a factor 6 lower than the threshold measured at lower $\chi$. These results show that the engineering of molecules that are less prone to quenching effects at high concentrations might constitute an effective route for decreasing the excitation thresholds in random arrays of complex fiber materials.

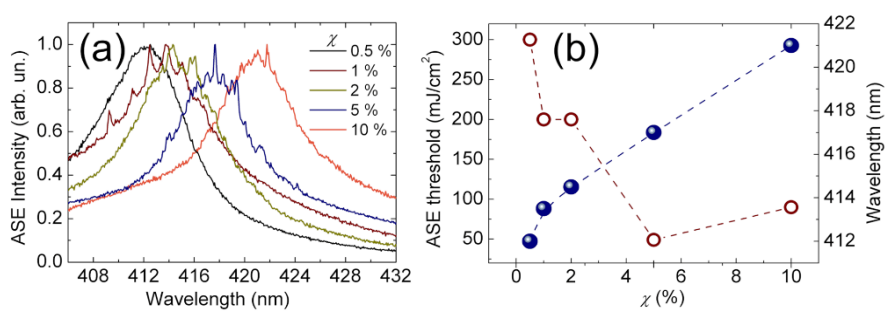

**Figure 7.** (a) ASE emission spectra from randomly oriented FCF-doped PS fibers as a function of the dye concentration. (b) Excitation threshold (empty circle, left vertical scale) and peak wavelength (full circle, right vertical scale) *vs.* FCF weight ratio for randomly oriented fibers. The peak wavelength is obtained by the maximum values of the normalized intensity shown in (a). Dashed lines are guides for the eyes.

## 3. Experimental

*Electrospinning*

Polymeric solutions used for the electrospinning of both uniaxially and randomly oriented fibers are prepared by dissolving 300 mg mL$^{-1}$ of PS in CHCl$_3$, and finally adding the FCF dye to the polymer solution at different weight ratios: 0.5%, 1%, 2%, 5% and 10% (w/w; dye/PS). The samples with TiO$_2$ nanoparticles are made by adding the particles (1% w/w; TiO$_2$/PS) to a solution of FCF with 5% w/w of dye with respect to the polymer matrix. Before electrospinning, the solutions are stirred for 12 h at room temperature for complete dissolution of the polymer and dye powder. In order to realize randomly oriented fibers, each solution is delivered at a constant flow rate of 1 mL h$^{-1}$ (Harvard Apparatus syringe pump, Holliston, MA) to a metal capillary (21 G) connected to a high-voltage power supply (EL60R0.6-22, Glassman High Voltage, High Bridge, NJ). Upon applying a high voltage (13 kV), a fluid jet is extruded through the capillary. As the jet accelerates towards a grounded plate collector, polymer fibers are deposited on a 10 × 10 cm$^2$ metallic collector placed at 10 cm from the needle. To prepare aligned fibers, the solution is delivered at flow rate of 1 mL h$^{-1}$ through a 21 G metallic needle and a metallic disk (8 cm diameter and 1 cm thickness) rotating at 3500 rpm (RT collector, Linari Engineering, Pisa, Italy) and positioned at 10 cm from the needle is exploited as the collector, with an applied high voltage of 13 kV. For both uniaxially and randomly oriented fibers, quartz substrates (1 × 1 cm$^2$ and thickness of 1 mm) are positioned onto the metallic collectors, allowing fibers to be deposited on the surface of the quartz. While the variation of the FCF content does not vary appreciably the morphology of the electrospun fibers, the incorporation of TiO$_2$ nanoparticles slightly affects the fiber shape in the case of significant TiO$_2$ agglomeration (Fig. S3, SI).

The morphology of electrospun fibers is inspected by SEM (FEI, Hillsboro, OR) and by STEM (FEI, Hillsboro, OR) after chromium coating. The elemental analysis is performed by EDX measurements. The diameter of the fiber is measured from SEM micrographs using an image analysis software (ImageJ).

*Absorption and ASE*

Absorption and emission measurements are preliminary performed on spin-coated films of PS doped with FCF, deposited on quartz substrates, by a spectrophotometer (Varian Cary 300 Scan). For ASE characterization, samples of fibers deposited on quartz substrates in aligned and random configurations are excited under vacuum conditions by the third harmonic of a pulsed Nd:YAG laser ($\lambda_{exc}$ = 355 nm, repetition rate = 10 Hz, pulse duration about 10 ns). All the measurements are performed in a side-pump configuration in which the excitation spot is focused on the samples in a stripe shape (maximum length = 4 mm, width = 50 μm), and with an end placed on the emitting edge of the samples. The signal is collected by a lens system and analyzed by a spectrograph. For ASE threshold measurements, the area of the pumping stripe is kept fixed, while varying the excitation fluence. For the gain spectrum characterization, the stripe length is varied, keeping fixed the excitation fluence at a value between the threshold and the saturation. Optical losses are studied by varying the distance of the stripe (at fixed dimension and fluence) from the edge of the sample.

## 4. Conclusions

We have studied electrospun fibers made of PS doped with the synthesized dye, FCF, in terms of optical gain as a function of both the FCF concentration and the fiber configuration (uniaxially aligned arrays or randomly oriented filaments). The FCF dye turns out to be marginally affected by quenching effects upon increasing the concentration up to 10% of FCF/PS weight ratio in electrospun fibers. The gain parameters of uniaxially aligned fibers reach optimal values with FCF concentration in the interval 2–5%, with the additional property of remarkable gain tunability. In randomly oriented fibers, the possibility to increase the dye amount without detrimental concentration-induced quenching effects leads to significantly reduced excitation threshold for the broadband lasing emission. Such results might be relevant for the exploitation of these laser sources in imaging, spectroscopy and optical sensing, where their RL properties in combination with other advantageous features of electrospun fibers, such as the high surface-to-volume ratio, might be put in synergy for enhancing device operational ranges and sensitivity.

**Acknowledgements**

A. Camposeo acknowledges funding from the Italian Minister of University and Research PRIN 201795SBA3. G. Morello acknowledges funding from POR Puglia 2014/2020, "Research for Innovation – REFIN" F9813165.

Supplementary Information

**Tuneable optical gain and broadband lasing driven in electrospun polymer fibers by high dye concentration**

*Giovanni Morello, Maria Moffa, Martina Montinaro, Annachiara Albanese, Karolis Kazlauskas, Saulius Jursenas, Ausra Tomkeviciene, Juozas V. Grazulevicius, Andrea Camposeo[*] and Dario Pisignano[*]*

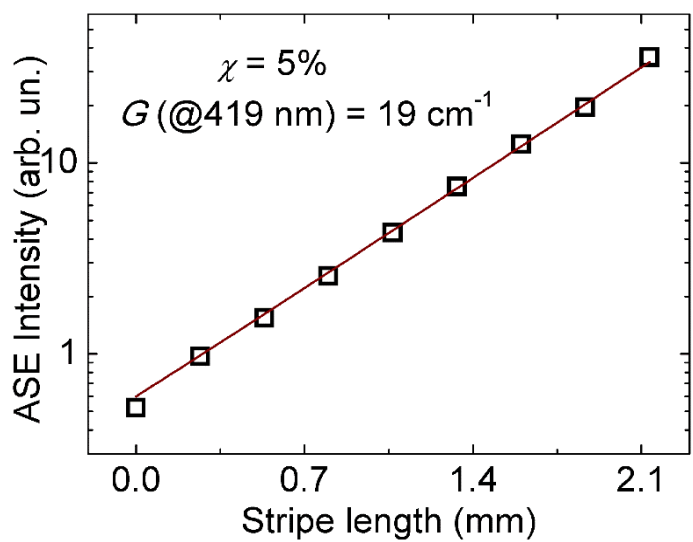

**Figure S1**. Dependence of the ASE intensity ($\chi$ = 5%, wavelength = 419 nm) on the excitation stripe length (squares) and corresponding best fit to the Eq. (1) (continuous line).

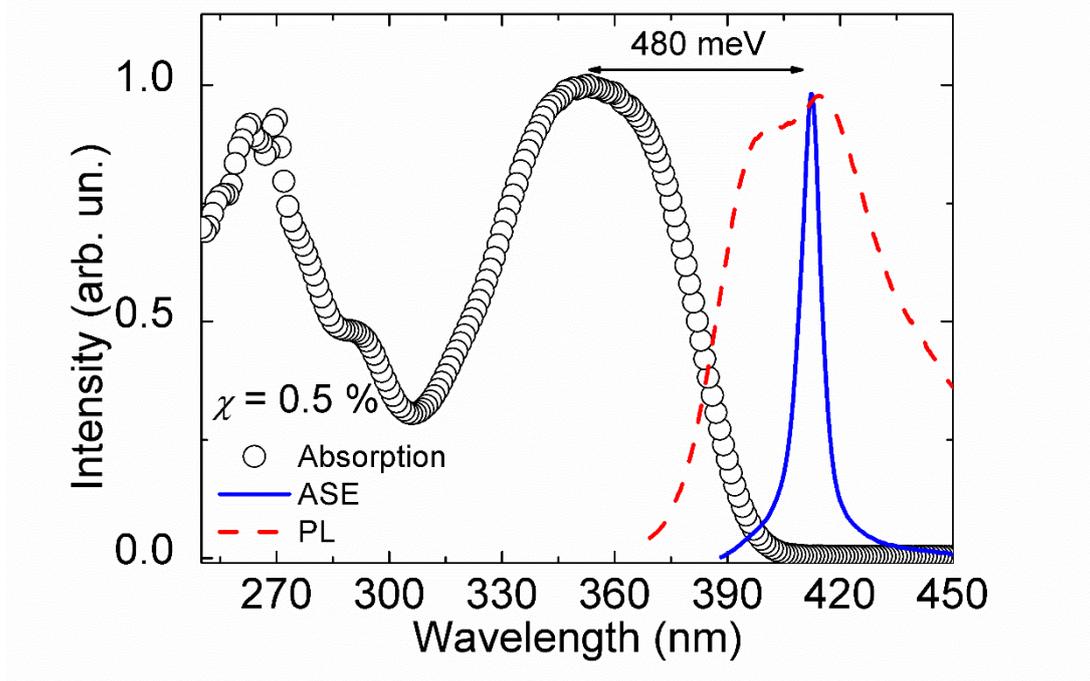

**Figure S2**. Normalized absorption (empty circles), photoluminescence (PL, red dashed line) and amplified spontaneous emission (ASE, blue continuous line) spectra of spin- coated films of FCF. The shift between the peak wavelengths of the absorption and ASE spectra is 480 meV.

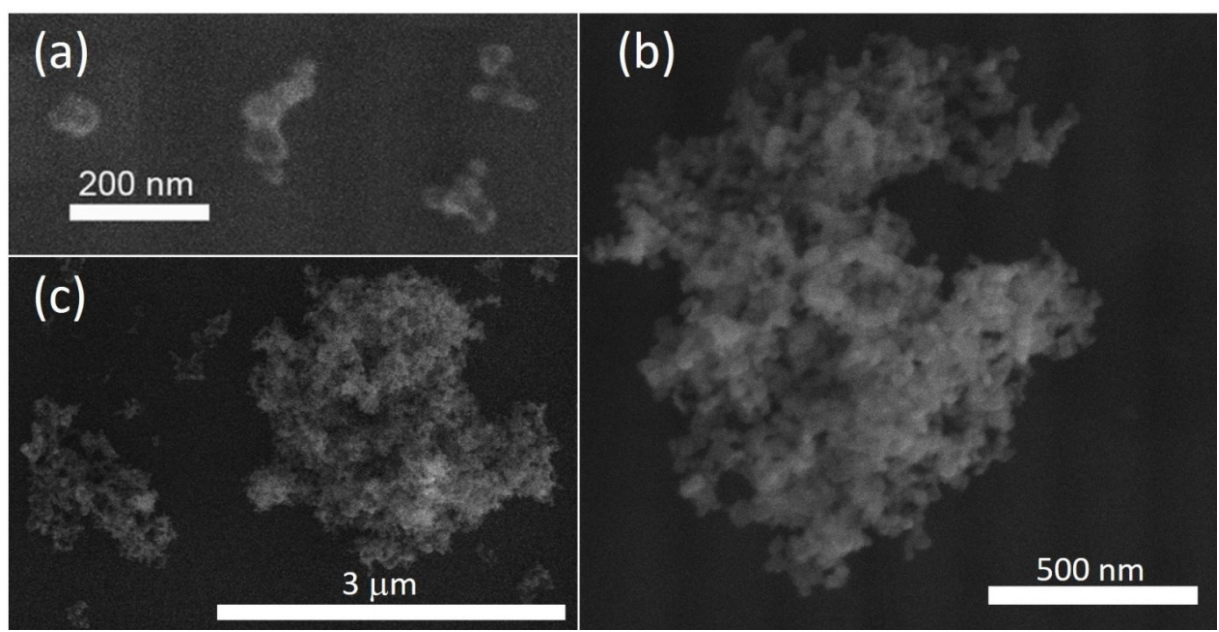

**Figure S3**. SEM micrographs of a few $TiO_2$ nanoparticles (a) and clusters (b,c).

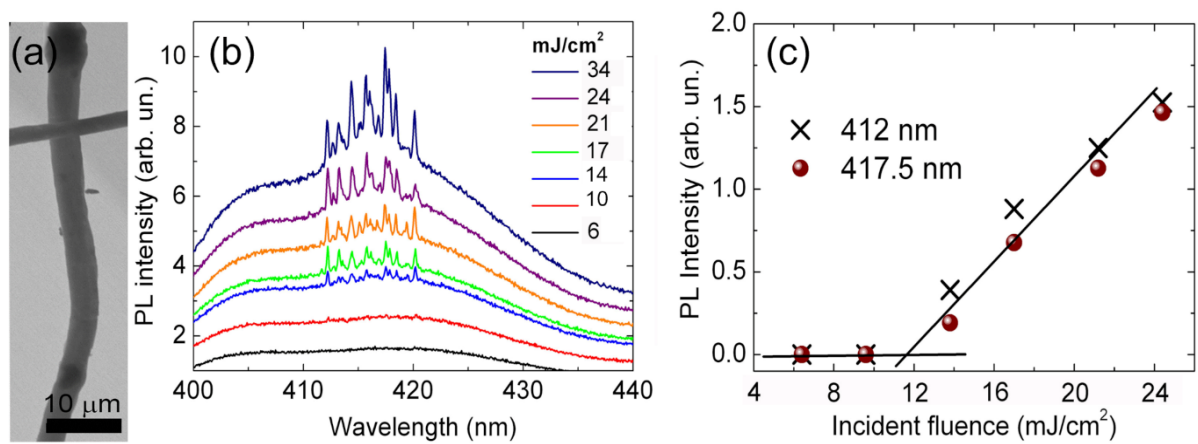

**Figure S4**. Morphological, compositional and RL analysis of randomly oriented fibers with $\chi = 5\%$ and embedding $TiO_2$ nanoparticles. (a) Scanning transmission electron microscopy (STEM) micrographs of electrospun fibers doped with $TiO_2$. (b) Emission spectra measured at different incident fluences. (c) Plot of the emission intensity *vs.* incident fluence of two representative RL peaks.

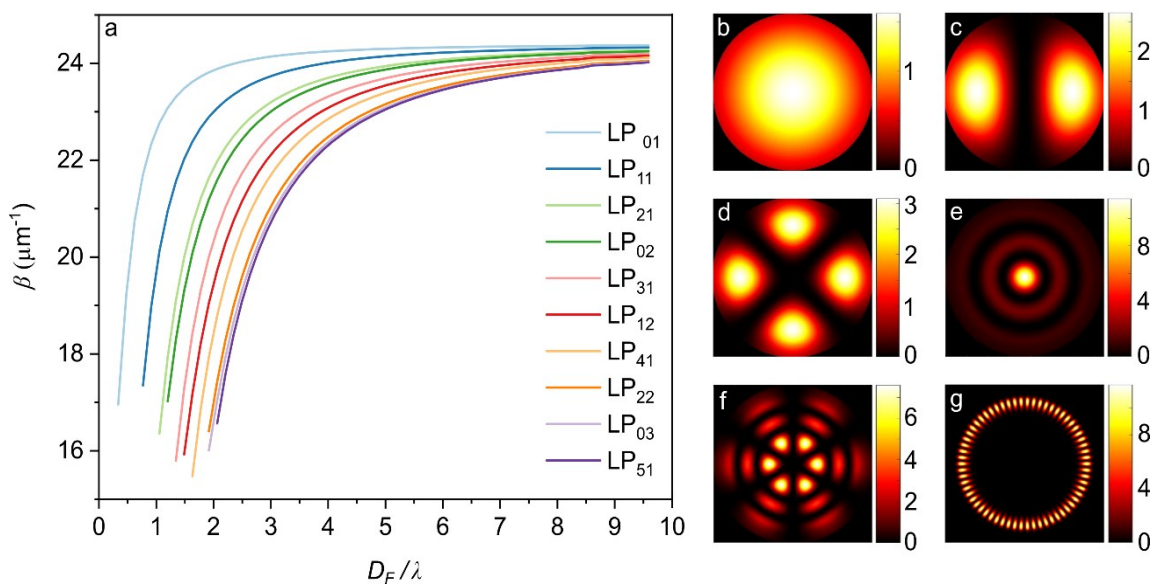

**Figure S5**. Calculated LP modes within a PS fiber. (a) Propagation constant, $\beta$, of the first ten LP modes as a function of the parameter $D_F/\lambda$, where $D_F$ is the diameter of the fiber and $\lambda = 417$ nm (refractive index $n \cong 1.62$ [S1]). (b-g) Examples of intensity spatial profiles for mode (b) $LP_{01}$, (c) $LP_{11}$, (d) $LP_{21}$, (e) $LP_{03}$, (f) $LP_{33}$ and (g) $LP_{31\,3}$. The values of $\beta$ and the modes spatial profiles are calculated by using the package of Ref. [S2].

**Table S1**. Comparison of optical properties and dye concentration reported for various classes of doped polymer fibers (aligned systems).

| Dye/polymer | Net gain (cm$^{-1}$) | $\chi$[a] | $\lambda_{em}$[b] (nm) | Reference |
|---|---|---|---|---|
| 4- (dicyano-methylene)-2-tert-butyl-6(1,1,7,7-tetramethyljulolidyl-9- enyl)-4H- pyran /polystyrene | 31.4 | 2% | 605 | [S3] |
| 2,7-bis(9,9-diethylfluoren-2-yl)-9-(2-ethylhexyl)carbazole/polystyrene | 24 | 2% | 415 | This work |
| 4-(dicyanomethylene)-2-tert-butyl-6(1,1,7,7-tetramethyljulolidyl-9-enyl)-4H-pyran /polyvinylpyrrolidone | 23.75 | 2% | 670 | [S4] |
| 2-[2-[3-[[1,3-dihydro-1,1-dimethyl-3-(3-sulfopropyl)-2H-benz[e]indol-2-lidene]ethylidene]-2-[4-(ethoxycarbonyl)-1-piperazinyl]-1-cyclopenten-1-yl]ethenyl]-1,1-dimethyl-3-(3-sulfopropyl)-1H-benz[e]indolium hydroxide, inner salt, compound and N,N-diethylethanamine (1:1)/poly(methyl methacrylate) | 7 | 0.5-1% | 950 | [S5] |
| 2-[[2-[2-[4-(dimethylamino)phenyl]ethenyl]-6-methyl-4H-pyran-4-ylidene]methyl]-3-ethyl iodide/poly(methylmetacrylate) | 5.5 | 0.16% | 740 | [S6] |
| 4,4‴-bis[(2-butyloctyl)oxy]- 1,1′:4′,1″:4″,1‴-quaterphenyl/poly(methylmetacrylate) | 5.4 | 1% | 387 | [S7] |
| 5-chloro-2-[2- [3-[(5-chloro-3-ethyl-2(3H)-benzothiazol- ylidene)ethylidene]- 2-(diphenylamino)-1-cyclopenten-1-yl]ethenyl]-3-ethyl benzothiazolium perchlorate/ poly(methylmetacrylate) | 4.2 | 0.5% | 910 | [S6] |

[a] ($\chi$ is expressed as the weight ratio between the incorporated dye and the polymer matrix).
[b] ($\lambda_{em}$ is the wavelength of net gain).